\documentstyle[11pt]{article}

\title{\bf High-colour pomerons  with a
running coupling constant in the Hartree-Fock approximation}
\author{M.A.Braun \thanks{On leave of absence from the Department of
high-energy physics,
 University of S. Petersburg, 198904 S. Petersburg, Russia}  \\
Department of Particle Physics, University of Santiago de Compostela,\\
15706 Santiago de Compostela, Spain}
  \date{November 1994}
\def\beq{\begin{equation}}
\def\eeq{\end{equation}}
\def\noi{\noindent}
\def\oq{\omega(q)}

\def\eq{\eta (q)}
\def\ea{\eta (q_{1})}
\def\eb{\eta (q_{2})}
\def\ec{\eta (q'_{1})}
\def\ed{\eta (q'_{2})}
\begin{document}
\maketitle
\medskip
\noi{\bf Abstract.} The Hartree-Fock approximation is applied to study the
"high-colour pomerons" in the system of many reggeized gluons with a running
QCD coupling constant. It is shown that, contrary to the fixed coupling case,
the high-colour pomerons result supercritical, although with a smaller
intercept than the multipomeron states.
\vspace{3 cm}

{\Large\bf US-FT/20-94}
\newpage

{\bf 1. Introduction.}
We have recently proposed a new method to incorporate the running QCD
coupling constant into the equation for two reggeized
gluons, based on  the so-called "bootstrap condition"  which requires that
in the gluon colour channel the solution  should be the reggeized gluon
itself [ 1 ]. As a result, the gluon  Regge
trajectory $\omega$ and its interaction $U$ are expressed via
a single function
$\eta$.  In the momentum space
\beq
\oq=-(3/2)\eq\int (d^{2}q_{1}/(2\pi)^{2})/\ea\eb
\eeq
and
\beq
U(q,q_{1},q'_{1})=-3(\ea/\ec+\eb/\ed)/\eta(q_{1}-q'_{1})+3\eq/\ec\ed
\eeq
In (2) $q$ is the total momentum of the two interacting gluons. In both
equations $q_{1}+q_{2}=q$. The fixed coupling constant BFKL pomeron [ 2 ]
corresponds to the choice $\eq=(2\pi/g^{2})q^{2}$. The running coupling
constant requires that  $\eq\simeq (2\pi/g^{2}(q))q^{2}$ for large $q$.
Parametrizing the confinement effects by a "gluon mass" $m$ we chose in
[ 1 ]
\beq
\eq=(b/2\pi)(q^{2}+m^{2})\ln ((q^{2}+m^{2})/\Lambda^{2})
\eeq
where $b=(11-(2/3)N_{F})/4$ and $\Lambda\simeq 200\ MeV$ is the standard QCD
parameter. Variational calculations with (1)-(3) gave the pomeron intercept
weakly dependent on $m$ and its slope rapidly falling with $m$ [ 1 ] .
With the slope of the order 0.2 $GeV^{-2}$, as favoured by the experimental
data, the gluon mass results in the region 0.6$\div$0.8 $GeV$ and the
intercept is of the order 0.35.

Thus nothing very spectacular ocurs for the perturbative pomeron upon the
introduction of the running coupling, except, of course, a nontrivial slope,
which, after all, has to be expected, since the BFKL pomeron does not contain
parameters with a dimension (see  [ 3 ]). In this note we study the effect
of the running coupling constant on the spectrum of the system of many
reggeized gluons. This system is described by the
Bartels-Kwiecinsky-Praszalowicz equation [ 4 ], which is nothing but a
Schroedinger equation for many gluons in the transverse space in which the
gluon trajectory serves as a kinetic energy (with a minus sign) and $U$
provides for the interaction. Exact solutions of this equation are not known
for the number of gluons $n\geq 2$. However there is an intense activity
around this problem in the limit of large number of colours $N_{c}$, when,
with a fixed coupling constant, the system turns out to be integrable [ 5,6 ].
For the realistic case $N_{c}=3$ and fixed coupling, the odderon case (n=3)
was studied in the variational approach in [ 7 ] with the result that the
odderon has an intercept greater than 1 although smaller than the pomeron.
Also for fixed coupling we studied the
$n$-gluon problem in the Hartree-Fock approximation (HFA) [ 8 ]. Looking
for solutions different from the multipomeronic ones we tried to evaluate
the intercept of states symmetric both in colour and ordinary space
variables. The result proved to be disappointing: in the HFA such states
possess intercepts lower than unity and going to the left linearly in $n$.
Such states cannot contribute at high energies.

The purpose of this note is to report that the introduction of the running
coupling changes the picture radically. Now the "energy" per gluon results
negative, which means that the intercepts of symmetric multigluon states are
above unity and go the right linearly with $n$. They still stay  lower
than those of the multipomeronic states which also exist in the $n$-gluon
system. However at high energies the symmetric multigluon states
("high-coloured pomerons") do not die out and give a nontrivial factor to
the standard contribution coming from multipomeron states. We recall that
such high-coloured pomerons were introduced  as high coloured strings in a
more phenomenological treatment in [ 9 ]. They seem to lead to a better
agreement with experiment for strange baryon production [ 10 ] and could be
felt in long-order correlations  [ 11 ]. The results of the present study
give some theoretical foundation for the existence of high-coloured strings.
\vspace{0.5 cm}

{\bf 2. The HFA for reggeized gluons.} After the change of the wave function
$\psi\rightarrow\prod\eta_{i}\psi$
where $\eta_{i}=\eta (q_{i})$, the equation for $n$ reggeized gluons aquires
the form
\beq H\psi=E\prod_{i=1}^{n}\eta_{i}^{2}\psi
\eeq
 The Hamiltonian $H$ is
given by a sum of pair terms
\beq H=-(1/6)\sum_{i<k}T_{i}T_{k}H_{ik}
\eeq Here $T_{i}$ is the colour vector of the $i$th gluon. In a colourless
state $\sum_{i=1}^{n}T_{i}=0$.
The pair Hamiltonian $H_{ik}$ is
\beq
H_{ik}=\prod_{j=1,j\neq
i,k}^{n}\eta_{j}(-\eta_{i}\eta_{k}(\omega_{i}+\omega_{k})+\eta_{i}V_{ik}
\eta_{k}+
\eta_{k}V_{ik}\eta_{i}+Q_{ik})\eeq
where $\omega_{i}=\omega(q_{i})$; $V_{ik}$ is a local potential acting
between the gluons $i$ and $k$ with the Fourier transform $\eq$:
\beq
V_{ik}=V(r_{ik})=\int (d^{2}q/4\pi^{2})\exp (iqr_{ik})/\eq
\eeq
and $Q_{ik}$ is a separable interaction whose action on the wave function is
defined as
\beq
Q_{ik}\psi(...q_{i},...q_{k},...)=\eta(q_{i}+q_{k})\int (d^{2}q/4\pi^{2})
\psi(...q'_{i},...q'_{k},...)
\eeq
with $q_{i}+q_{k}=q'_{i}+q'_{k}$ and $q'_{i}-q'_{k}=2q$. The intercept is
given by the energy with a minus sign
\beq
E=1-j=-\Delta
\eeq
so that the rightmost singularity in the complex angular momentum plane
corresponds to the ground state.
The solution of (4) may  be found by a variational approach, as
realizing the minimum value of the functional
\beq
\Phi=\int\prod_{i=1}^{n} d^{2}r_{i}\psi^{\ast}H\psi
\eeq with the normalization condition
\beq
\int \psi^{\ast} \prod_{i=1}^{n}d^{2}r_{i}\eta_{i}\psi=1
\eeq

Eq. (4) mixes the colour and space variables. So its solution cannot be
generally presented as a product of a colour wave function and a spatial
one, both symmetric in their respective variables, except in some special
cases (n=2,3 and the limit
$N_{c}\rightarrow\infty)$. In the HFA, however, one assumes that each gluon
is moving in some average field, which in a colourless state cannot depend
on colour. So in the HFA the colour and spatial variables decouple and the
total wave function may be taken as a product of two symmetric functions in
colour and in space. This allows to simplify the variational treatment
considerably. Indeed, in a colourless state with a symmetric colour wave
function
\beq <T_{i}T_{k}>=(2/n(n-1))\sum_{i<k}T_{i}T_{k}=-3/(n-1)
\eeq  where  $T^{2}=3$ has been used. With this result and the symmetry of
the spatial wave function we find that the energy can be found from the
minimum  of a simpler functional in only spatial variables
\beq {\cal E}=(1/2)\int\prod_{i=1}^{n}d^{2}r_{i}\psi^{\ast}H_{12}\psi
\eeq where the Hamiltonian $H_{12}$ is defined by (6) with $ik=12$ and the
function $\psi$ should satisfy (11). The energy of the whole system of $n$
gluons is determinned by the minimal value  $\epsilon$ of $\cal E$  according
to \beq E_{n}=(1/2)n\epsilon_{n}
\eeq Actually the operator in $\cal E$ does not depend on $n$. The dependence
on
$n$ enters only from extra arguments in $\psi$ through the requirement of the
symmetry in all arguments.

In the HFA one takes for $\psi$ a product
\beq
\psi=\prod_{i=1}^{n}\psi(r_{i})
\eeq
with the normalization
\beq
\int d^{2}r\psi^{\ast}\eta\psi =1
\eeq
Then $\cal E$ reduces to a functional which involves the gluons 1 and 2
exclusively and does not depend on $n$ altogether:
\beq {\cal E}=(1/2)\int d^{2}r_{1}d^{2}r_{2}\psi^{\ast}(r_{1})
\psi^{\ast}(r_{2})\tilde{H}_{12}\psi (r_{1})\psi(r_{2})
\eeq
where $\tilde{H}_{12}$ is given by (6) without the first factor (the product
of all $\eta_{i}$ except $i=1,2$).
As a result in the HFA $\epsilon$ does not depend on $n$ and the total
energy is linear in the number of gluons
\beq E_{n}^{HF}=(1/2)n\epsilon
\eeq
If $\epsilon<0$ all high-coloured pomerons result supercritical in this
approximation, the intercept growing linearly with $n$. If $\epsilon>0$ the
they are all subcritical and the intercept moves to the left linearly in
$n$.

Calculations carried out in [ 1 ] for a fixed coupling constant gave
$\epsilon\simeq (3\alpha_{s}/\pi)0.959$, so that all high-coloured states
turned out subcritical and irrelevant at high-energies (in the
HFA).\vspace{0.5 cm}

{\bf 3. Calculations with the running coupling constant.}
With the running coupling we choose $\eq$ according to (3) with the gluon
mass $m$ as a parameter. From the comparison
of the pomeron slope to  experimental data its value can be taken to lie in
the range 0.6 $\div$ 0.8 $GeV$. The functional $\cal E$ was studied on the
gluonic functions $\psi(r^{2})$ which were taken to be a sum of Gaussians
\beq
\psi(r^{2})=\sum_{k=1}^{N}c_{k}\exp (-\alpha_{k}r^{2}/2)
\eeq
with $\alpha_{k+1}=c\alpha_{k}$.  The calculation is
straightforward, although very time consuming due to a large number of
successive integrations (up to 6) and the necessity of a high precision
because of cancellations between the kinetic and potential energies. The
energy seems to go down with the rise of $\alpha_{N}$ but the calculational
error rapidly grows with it. For that
reason we had to limit ourselves with $N=3$ and have chosen $c=3$. In the
end the variaitional parameters are $\alpha_{1}$ and $c_{k}$.

The results of the calculations for $m$ in the region 0.28 $\div$ 2.0 $GeV$
are presented in the Table. Intercepts per gluon $\delta=\Delta/n$ together
with the optimal values for $\alpha_{1}$ and $c_{k}$ are given in the Table.
To see the convergence in $N$ the results for $N=2,3$ are shown.
One observes that $\delta$ is negative for values of $m$ close to the value
of $\Lambda=0.2\ GeV$ in agreement with the fact that in the limit
$m\rightarrow\Lambda$ our equation goes over into the BFKL one [ 1 ].
However with the rise of $m$ the intercept goes up rather steeply and
becomes positive already at $m=0.4 \ GeV$. It then stays practically
independent of $m$, of the order $\delta= 0.08-0.09$, in the region $m=0.6\
\div\ 2.0\ GeV$ slightly falling at the end.
 With the realistic values of $m$ of the order $0.6\ -\ 0.8\ GeV$, this
means that the high-coloured pomerons are supercritical and their intercept
goes up linearly with $n$. To compare, we  present the intercepts per
gluon
$\delta_{P}$ of the $n/2$ pomeron state
 which is also present in the multigluon system and corresponds to a
multipomeron cut in the old Regge-Pomeron theory
(one half of the pomeron intercept calculated in [ 1 ] for different values of
$m$). We observe that  the intercepts of the high-colour pomerons are on the
average nearly twice smaller than those of the mutipomeronic states in the
studied range of $m$.
\vspace{0.5 cm}

{\bf 4. Discussion.} Although the obtained results  are rather
direct and transparent by themselves, they should be taken with some caution.
First, the fact that the "energy" of the new high-colour pomeron lies
 above the energy of  multipomeron states means
that it is unstable and decays into pomerons with less colour, eventually
into ordinary pomerons. It is difficult to estimate its lifetime. It may be
not so small because of the difference in the space structure of the new
pomeron, with a wave function symmetric in all gluons and rapidly falling at
large distances, and a mutipomeronic state, which is not symmetric in
gluons and has a wave fucnction which does not fall as the pomerons go
apart. Still the decays may essentially change the role of the new states at
high energies.

Another suspicious point is that the parameter $\alpha_{1}$ in the
optimal wave function turns out to be of the same order as the gluon mass
squared.
 This parameter, in fact, determines the average momentum squared. Thus
the averaged momenta of our solutions are of the order $m$ and  lie in the
nonperturbative region. In other words, our new pomerons have a
transverse dimension comparable to that of a hadron. That
means that they should be  sensitive to what we assume about the behaviour of
$\eq$ at low $q$. Negative $\epsilon$'s
we have obtained may be a result of our particular choice of $\eq$ and may
turn into positive ones for a different choice (say, $\eq =(2\pi/b)q^{2}\ln
((q^{2}+m^{2})/\Lambda^{2})$). This does not make our results invalid, since,
after all, the behaviour of $\eq$ at small $q$ is completely unknown and out
of control. An experimental confirmation of the existence of high-coloured
strings would be a decisive argument in this respect.\vspace{0.5 cm}

 {\bf 5. Acknowledgments.}
The author is deeply indebted to Prof. C.Pajares for attracting  attention
to the problem and  fruitful and constructive discussions. He also expresses
his gratitude to the General Direction of the Scientific and Technical
Investigation (DGICYT) of Spain for financial support.\vspace{0.5 cm}

{\bf 6. References.}

\noi 1. M.A.Braun, Univ. of Santiago preprints US-FT/11-94, US-FT/17-94\\
\noi 2. E.A.Kuraev, L.N.Lipatov and V.S.Fadin, Sov. Phys. JETP {\bf 44}
(1976) 433; {\bf 45} (1977) 199\\ Ya.Ya. Balitzky and L.N.Lipatov, Sov.Phys.
J.Nucl. Phys. {\bf 28} (1978) 822\\
\noi 3. N.N.Nikolaev, B.G.Zakharov an V.R.Zoller, ITEP preprint ITEP-74-94
(1994)\\
\noi 4. J.Bartels, Nucl.Phys.{\bf B175} (1980) 365\\ J. Kwiecinsky and
M.Praszalowicz, Phys. Lett. {\bf B94}(1980) 413\\
\noi 5. L.N.Lipatov, JETP Lett.{\bf 59} (1994) 596\\
\noi 6. L.D.Faddeev and G.P.Korchemsky, preprint ITP-SP-14 (
hep-th/9404173).\\
\noi 7. P.Gauron, L.N.Lipatov and B.Nicolescu, Z.Phys. {\bf C63} (1994)
253.\\
\noi 8. M.A.Braun. Phys. Lett. {\bf B337} (1994) 354\\
\noi 9. M.A.Braun and C.Pajares, Phys. Lett. {\bf B287} (1992) 154; Nucl.
Phys. {\bf B390} (1993) 542,559\\
\noi 10.  N.Armesto, M.A.Braun, E.G.Ferreiro and C.Pajares, Univ. of Santiago
preprint US-FT/16-94 (to be published in Phys. Lett.{\bf B})\\
\noi 11. N.S.Amelin, N.Armesto, M.A.Braun, E.G.Ferreiro and C.Pajares,
Phys. Rev. Lett. {\bf 73} (1994) 2813\\

\newpage \vspace*{3 cm}
{\large\bf Table. High-colour pomeron intercepts per gluon}
\vspace {1 cm}
\begin{center}
\begin{tabular}{|c|c|c|c|c|c|}\hline
$m$&$\delta (N=2)$&$\delta (N=3)$&$\delta_{P}$& $\alpha_{1}$
& $c_{k}(N=3)$\\\hline
0.28  &-0.0281& -0.0212&  0.290& 0.0590& 0.957,-0.107,0.269\\\hline
0.4  & 0.0427&  0.0486&  0.225&  0.176& 0.940,-0.146,0.309\\\hline
0.6  & 0.0752&  0.0820&  0.186&  0.324& 0.910,-0.208,0.358\\\hline
0.8  & 0.0816&  0.0896&  0.168&  0.544& 0.888,-0.358,0.381\\\hline
1.0  & 0.0821&  0.0907&  0.156&  0.825& 0.871,-0.286,0.400\\\hline
1.41 & 0.0792&  0.0877&  0.140&    1.6& 0.854,-0.304,0.423\\\hline
2.0  & 0.0737&  0.0821&  0.128&    3.2& 0.842,-0.324,0.432\\\hline
\end{tabular}
\end{center}
\vspace{1 cm}

{\large\bf Table captions}
\vspace{0.3 cm}

The first column gives values of the gluon mass  in $GeV$.
In the second and third columns  the corresponding calculated high-colour
pomeron intercepts per gluon are presented for $N=2$ and $N=3$ respectively.
In the fourth column the intercepts per gluon of the multipomeronic state,
taken from [ 1 ], are shown for comparison. The fifth column shows the
optimal values of the variational parameter
$\alpha_{1}$ in $GeV^{2}$. In the last column the optimal values for $c_{k}$
$k=1,...N$, are presented for  $N=3$.
 The QCD parameter $\Lambda$ has been taken to be $0.2\ GeV$ for
three flavours ($b=9/4$). The calculational error for the intercept is
estimated to be of the order $\pm 2.10^{-3}$ for all values of $m$ except
$m=0.28\ GeV$ when it may rise up to $\pm 4.10^{-3}$.

 \end{document}